# First Demonstration of β-Ga$_2$O$_3$ Optical Waveguides and the Analysis of Their Propagation Losses in the UV–Visible Spectra

Jingan Zhou[1], Hong Chen[1], Houqiang Fu[1], Kai Fu[1], Xuguang Deng[1], Xuanqi Huang[1], Tsung-Han Yang[1], Jossue A. Montes[1], Chen Yang[1], Xin Qi[1], Baoshun Zhang[2], Xiaodong Zhang[2, b)], Yuji Zhao[1, a)]

[1]*School of Electrical, Computer and Energy Engineering, Arizona State University, Tempe, AZ 85287, USA*

[2]*Key Laboratory of Nanodevices and Applications, Suzhou Institute of Nano-tech and Nano-bionics, CAS, Suzhou 215123, China*

This paper reports the first demonstration of beta-phase gallium oxide (β-Ga$_2$O$_3$) as optical waveguides on sapphire substrates grown by metal organic chemical vapor deposition (MOCVD). The propagation losses from visible to ultraviolet spectra were comprehensively studied. By optimizing the fabrication processes, minimum propagation loss was identified to be 3.7 dB/cm at the wavelength of 810 nm, which is comparable to other wide bandgap materials within III-N family (GaN, AlN). To further reveal the underlying loss mechanisms, several physical mechanisms such as two-photon absorption, sidewall scattering, top surface scattering, and bulk scattering were taken into consideration. The results obtained from this work suggest that β-Ga$_2$O$_3$ is promising for ultraviolet–visible spectrum integrated photonic applications.

Photonic integrated circuits (PICs) have shown excellent performance in high-speed signal transmission and processing compared with traditional discrete optical components. In the past few years, silicon photonics has attracted considerable attention as an excellent candidate for PICs and has exhibited good compatibility with the mature CMOS processes.[1,2] However, the narrow bandgap (1.1 eV) of silicon restricts the light transmission to wavelengths longer than 1130 nm, which hinders its applications in frequency metrology,[3,4] on-chip mode-locking,[5] visible light communications,[6] spectroscopy,[7] and biosensing,[8] where the transmission of ultraviolet (UV)–visible spectra is required. To extend the working wavelengths into the visible spectral, other wide-bandgap materials have been investigated, such as silicon nitrite with a bandgap of 5.0 eV at 300 K.[9] The disadvantage of silicon nitrite lies in the strong material absorption due to its N-H bond concentrations[10], which limits low-loss applications in the UV–visible spectral. The ideal candidates are expected to have smaller photon absorption. To tackle these problems, we proposed β-$Ga_2O_3$ as a new material platform for PIC applications in the UV–visible spectral. The wide bandgap of β-$Ga_2O_3$ (4.8 eV)[11] provides broadband transparency. Its small two-photon coefficient ($β_{TPA}$= 0.6 cm/GW at 400 nm)[12] allows weak photon absorption and two-photon effects compared with other III-V materials with narrow band gaps. Furthermore, it also has a small lattice mismatch with III-N material system,[13] which is beneficial to active integration of lasers and detectors. Therefore, β-$Ga_2O_3$ is promising for integrated photonic applications in the UV–visible spectral with low propagation loss.

Recently, several growth techniques were proposed for β-$Ga_2O_3$ thin film on sapphire substrates,[14] such as chemical vapor deposition (CVD),[15] metal organic chemical vapor deposition (MOCVD),[16] atom layer deposition (ALD),[17] and molecular beam epitaxy (MBE).[18] Among them, MOCVD growth method has been widely used for massive productions due to its high crystalline quality, fast growth rate, and good control on thickness and doping. In this work, the β-$Ga_2O_3$ films were grown on c-plane sapphire substrates by MOCVD (NIPPON SANSO reactor). High-purity $O_2$ gas and triethylgallium (TMGa) were utilized as oxygen and gallium sources, respectively. High-purity $N_2$ was used as the carrier gas for TMGa. The β-Ga2O3 film was grown at 750 °C and 760 Torr. The thickness of the β-Ga2O3 films was 1 μm measured by ellipsometer. Additionally, atom force microscope (AFM) was used to evaluate its surface morphology and roughness. The root-mean-square roughness was 8.4 nm in a scanning area of 5 μm ×5 μm.

Figure 1 shows the fabrication process of β-$Ga_2O_3$ waveguides. First, 1 μm of $SiO_2$ film and 80 nm of chromium (Cr) film were deposited on β-$Ga_2O_3$/$Al_2O_3$ sample by plasma-enhanced chemical vapor deposition (PECVD) and e-beam evaporation, respectively. Then a layer of 290 nm negative photoresist (maD-2403) was covered onto the Cr layer by spin-coating. The Cr layer served as the hardmask during $SiO_2$ etching and a conductive layer for electron beam lithography (EBL). After the EBL exposure and development in ma-D 525 developer, the waveguide patterns were defined on the sample. The Cr layer was etched by chlorine based inductively coupled plasma (ICP), and the $SiO_2$ layer was etched by fluorine based reactive ion etching (RIE). Then, the exposed β-$Ga_2O_3$ layer was removed by chlorine based ICP.[19] For the purpose of device protection and lower scattering loss, 2 μm PECVD $SiO_2$ was deposited on the sample to cover the waveguides.

Finally, the cross sections of the waveguides were exposed with a diamond blade cut, and diamond grind papers were employed to polish the cross sections to 0.1 μm grade, which enhanced the light coupling efficiency.

Figure 2(a) presents the propagation loss measurement system adopted in the work. Four different incident light wavelengths were chosen in the UV–visible spectral: 810 nm, 633 nm, 526 nm and 400 nm. The 810 nm wavelength was provided by a Ti:Sapphire laser with 100 fs pulse width and 82 MHz repetition rate. The latter three wavelengths were provided by three continuous-wave (CW) diode lasers. The waveguide mode was kept in fundamental TM polarization within this study. The incident light was focused by the objective lens and then coupled into the β-Ga$_2$O$_3$ waveguide sample, which was mounted on a XYZ three-dimensional translation stage for precision alignment. When the guide mode was propagating along the waveguide, a linear CMOS camera was located above the sample to record the top scattered light, which can represent the light power transmitting in the waveguides after analysis in a computer. This method is simple, accurate and nondestructive, and has less requirements to coupling efficiency,[20] compared with other techniques.[21,22] Figure 2(b) provides a typical recorded image from a linear CMOS camera during measurement, and Figure 2(c) shows the extracted out-scattered light. The distance dependence of the scattered light intensity and can be expressed as:

$$I_{z_2} = I_{z_1} e^{-\alpha(z_2-z_1)} \quad (1)$$

where $z_1$ and $z_2$ represent two points along the waveguide, $I_{z_1}$ and $I_{z_2}$ are the corresponding light intensities. The total loss coefficients can be obtained from Eq. (2):

$$\gamma = -10(z_2 - z_1)^{-1} \log(I_{z_2}/I_{z_1}) \quad (2)$$

The waveguides had same heights of 1 μm and different widths varying from 0.5 to 1.75 μm. Figure 3(a) shows that the propagation losses increased towards shorter widths and wavelengths. The minimum loss of 3.7 dB/cm was achieved at 1.5 μm width and 810 nm wavelength. The transmission spectrum of β-Ga2O3 indicates a high transmission in the visible spectral with an absorption edge at around 260 nm (253 nm for $\vec{E}\|\vec{b}$ and 270 nm for $\vec{E}\|\vec{c}$),[23] which is far away from the visible spectral. It suggests that the intrinsic material absorption is negligible in UV–visible spectral. Therefore, in the next section, we focus on other mechanisms such as the nonlinear two-photon absorption (TPA) and scattering.

Two-photon absorption becomes significant when the photon energy reaches half of bandgap energy. To estimate the TPA loss, the TPA coefficients at different wavelengths were theoretically calculated using the following equations:[12,24]

$$\beta(\omega) = K \frac{\sqrt{E_p}}{n_0^2 E_g^3} F_2\left(\frac{\hbar\omega}{E_g}\right) \quad (3)$$

$$F_2(x) = \frac{(2x-1)^{3/2}}{(2x)^5} \quad (4)$$

where $E_g$ is the bandgap energy of β-$Ga_2O_3$ and $E_p = 2|\vec{p}_{vc}|^2/m_0$, which is obtained using $\vec{k}\cdot\vec{p}$ theory. $n_0$ is the refractive index of β-$Ga_2O_3$, $\omega$ is the light frequency, and $K$ is a material-independent constant, which has a value of $K = 1940$ in units such that $\beta$ is in cm/GW and $E_g$ and $E_p$ are in eV. Figure 4 (a) shows the wavelength dependence of TPA coefficient, which is in accord with experimental results[12]. When the incident light wavelength is longer than 510 nm, the energy of two photons isn't able to excite an electron to the conductive band, and the TPA coefficient is reduced to zero. Equation (5) gives the decay of light intensity $I$, where α is the linear loss coefficient resulted from scattering and material absorption, and $\beta$ is the TPA coefficient. The second term in equation (5) was used to calculate TPA losses. The waveguide light intensity profile $I(x,y)$ of fundamental TM mode was calculated with a finite-difference method, and then applied to equation (6) to calculate the absorbed light power in unit length $L_0$:[25]

$$dI/dz = -\alpha I - \beta I^2 \quad (5)$$

$$P = \int_0^{L_0} \int_{-\infty}^{+\infty} \int_{-\infty}^{+\infty} (-\beta I^2)\, dx\, dy\, dz \quad (6)$$

Figure 4 (b) shows the TPA losses of a waveguide of 1.0 µm width at CW TM mode at different wavelengths and input power. Before calculation, a coupling loss of 3 dB was applied to the input power when the laser light was injected into waveguides. The TPA loss was influenced greatly by the input power and it only became significant at high power and wavelengths shorter than 500 nm. Since the input power in the experiment system was limited under 100 mW, there was only an imperceptible TPA loss in this paper. However, it's also worth noting that if the waveguides work under pulsed laser, the TPA loss can become significant. Assuming the same input light power and coupling loss in this calculation, the maximum light intensity (~$10^{12}$ W/cm$^2$) of a pulsed laser (assuming pulse width of 400 ps and repetition rate of 10 kHz) is $10^4$ times larger than light intensity (~$10^8$ W/cm$^2$) of a CW laser. Because of the intensity dependence, the TPA loss will become one of the main sources of propagation loss under pulsed laser.

For the scattering loss, generally, three mechanisms are studied corresponding to three regions in waveguides: top surface, sidewall and bulk scattering. The top surface scattering and sidewall scattering are influenced by the roughness of top surface and sidewall respectively. Both of them are also governed by the refractive index contrast between waveguide material and cladding material. Higher refractive index contrast will cause higher scattering losses.[26,27] The top surface roughness was mainly determined by the MOCVD processes. The sidewall roughness was mainly caused by different etching rates of crystal grains with different crystalline orientations. Figure 5 and Table I show the etching recipe parameters and scanning electron microscope (SEM) images before and after optimization of ICP etching recipe. The original recipe was chosen according to the previous research of chlorine based ICP.[28] However, the high BCl$_3$ gas flow and ICP power caused the high plasma density, which resulted in extra etching perpendicular to sidewalls as it's identified in Figure 5(a).

To properly model the scattering losses, the volume current method (VCM) was applied.[29,30] The roughness was firstly decomposed into an array of unit non-idealities, and each non-ideality can be treated as "volume current sources". By integrating the far-field radiated power from the volume current source and applying proper array factors, the total loss in dB/cm can be estimated. Using the dyadic Green's function derived from previous work,[29,30] the electric field of a single dislocation can be calculated using equation (7):

$$\vec{E}(\vec{r_c}) = i\omega\mu \iiint \vec{G}(\vec{r_c}, \vec{r_c}') \vec{J}(\vec{r_c}') d\vec{V}' \qquad (7)$$

where $\vec{r_c}$ is the displacement vector of electric field, $\vec{r_c}'$ is the displacement vector of volume current source, $\mu$ is the permeability constant, $\omega$ is the angular frequency of the incident light, $\vec{G}(\vec{r_c}, \vec{r_c}')$ is the corresponding Green's function, and $\vec{J}(\vec{r_c}')$ is the volume current density. The far-field Poynting vector and the corresponding radiated light power can be obtained with equation (8) and (9):

$$\vec{S} = \vec{r}\frac{\omega n_1 k_0}{2\mu}|\vec{r} \times \vec{E}|^2 \qquad (8)$$

$$P = \oiint(\vec{S} \cdot \vec{r}) dA \qquad (9)$$

where $n_1$ is the refractive index of the waveguide material, and $k_0$ is the free-space wavenumber. After applying the array factors[29] to radiated light power of a unit non-ideality, the total radiated power per unit length of the waveguide can be computed with equation (10) and (11):

$$\tilde{R}(\Omega) = 2\sigma^2 L_c / (1 + L_c^2 \Omega^2) \tag{10}$$

$$P/2L = 2 \oiint (\vec{S} \cdot \vec{r}) \tilde{R}(\beta - n_c k_0 \cos\theta) \, dA \tag{11}$$

where $\sigma$ is the roughness, $L_c$ is the correlation length set as 100 nm for simplicity, $n_c$ is the refractive index of cladding material, $\beta$ is the modal propagation constant, and the integration in equation (11) was in polar coordinates.

With an estimation of 10 nm sidewall roughness, assuming TM mode operation and waveguides height of 1 μm, the scattering losses of sidewall and top surface can be calculated as shown in Figure 6. Due to the 'squeezed out' effect, waveguides with smaller dimension exhibit larger sidewall scattering loss, especially when the width is narrower than 0.75 μm. On the other hand, the scattering losses are also sensitive to wavelengths. The UV light exhibit stronger scattering loss comparing with red light. This wavelength dependence was one of the reasons to the dramatic increase of total losses at UV spectral. It should be noted that the refractive index of β-Ga$_2$O$_3$ (n~1.85) is relatively small comparing with other III-N materials such as GaN (n ~ 2.35) and AlN (n ~ 2.12), since scattering loss is proportional to the contrast of dielectric constant between core and cladding materials ($\Delta\varepsilon$), β-Ga$_2$O$_3$ is intrinsically less vulnerable to scattering losses and is ideal for high quality resonators in UV–visible spectrum wavelengths. The appropriate index of β-Ga$_2$O$_3$ also allows single mode operation in relative larger waveguide dimensions, which is promising for the applications in on-chip high speed optical interconnections.

During the MOCVD growth, due to the large lattice mismatch between β-Ga$_2$O$_3$ and sapphire, large density of defects is expected, such as grain boundaries and threading dislocations (typically $10^8$–$10^{10}$ cm$^{-2}$ defect density[31]). The bulk scattering losses induced by bulk defects can be calculated by removing the TPA losses, top surface scattering losses and sidewall scattering losses from the total losses. Table II provides the contributions from each loss mechanisms for three cases (width = 1.5 um, λ = 810 nm; width = 0.5 um, λ = 810 nm; width = 1.5 um, λ = 526 nm). The TPA isn't influential at these wavelengths because of the wide bandgap of β-Ga$_2$O$_3$. For case (1), the bulk scattering is responsible for the total loss due to the relatively large waveguide width. While for case (2), the sidewall scattering dominates due to the overlap between mode and sidewalls. Finally, for case (3), which has a shorter wavelength than case (1), significant increase in total loss can be observed, which is the result from the boost of bulk scattering. The bulk scattering possesses strong wavelength dependence. These analyses implied that the smooth sidewall roughness and better crystal quality were the limiting factors of waveguide performance.

Comparing with other researches on waveguides in UV–visible spectral using other material platforms, such as AlN (~2.36 dB/cm at 905 nm, ~8 dB/cm at 390 nm),[32,33] the β-$Ga_2O_3$ platform demonstrated here performs decently, and we foresee large space for improvements. In this work, the typical surface roughness is 8.4 nm, which is far from the typical optimized roughness value from MOCVD process (< 3 nm). The roughness of sidewall is expected to decrease after further optimizations on etching recipe. Bulk scattering can also be minimized by improving crystalline quality.

To conclude, this paper provides the first demonstration of β-$Ga_2O_3$ waveguides on sapphire substrate, and the minimum propagation loss of 3.7 dB/cm was achieved at the wavelength of 810 nm, which is decent for a wide variety of applications. With the theoretical calculations using VCM, the TPA loss, sidewall scattering loss, top surface scattering loss and bulk scattering loss were estimated. The TPA loss is negligible for low power operation. Sidewall scattering mechanism domains in narrow widths, and the bulk scattering mechanism domains in waveguide with larger dimensions. All these scattering mechanisms increase with decreasing wavelengths. This work provides new platform for on-chip high speed interconnections and UV–visible nonlinear optical applications.

**Acknowledgements**
Access to the ASU NanoFab was supported, in part, by NSF Contract No. ECCS1542160. We gratefully acknowledge the use of facilities within John M. Cowley Center for High Resolution Electron Microscopy (CHREM), Goldwater Materials Science Facility (GMSF), Ultra-Fast Laser Facility and Ning's Nanophotonics Lab in Arizona State University.

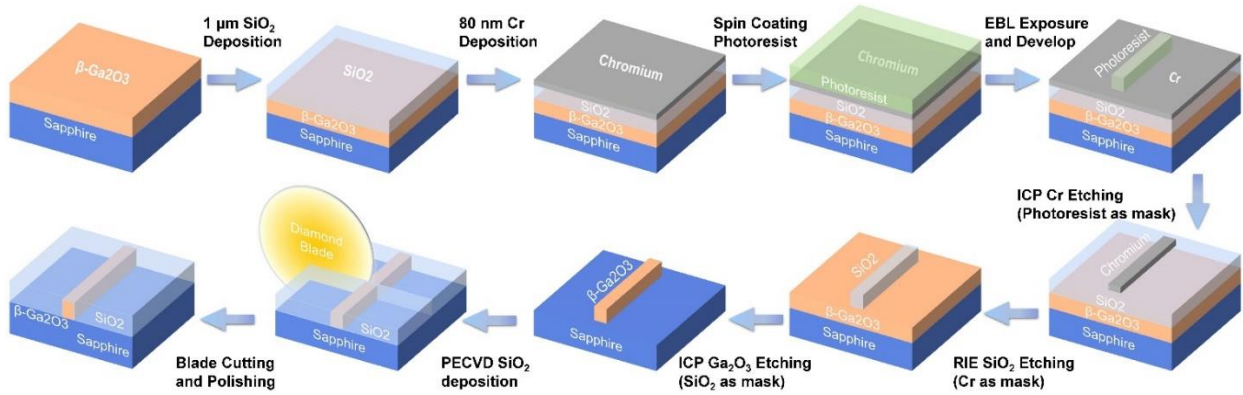

FIG. 1. Illustration of the fabrication processes of β-Ga$_2$O$_3$/Sapphire waveguides.

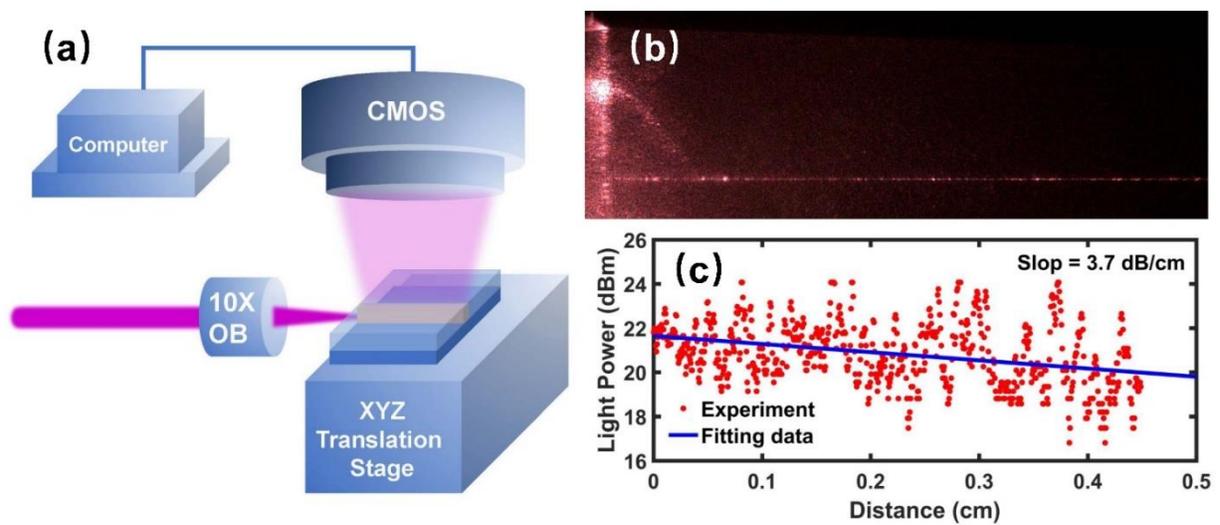

FIG. 2. (a) Schematic of measurement system. (b) Top image captured by the linear CMOS camera of a waveguide of 1.5 μm width at 810 nm wavelength. (c) Experimental data and regression analysis of 3.7 dB/cm loss.

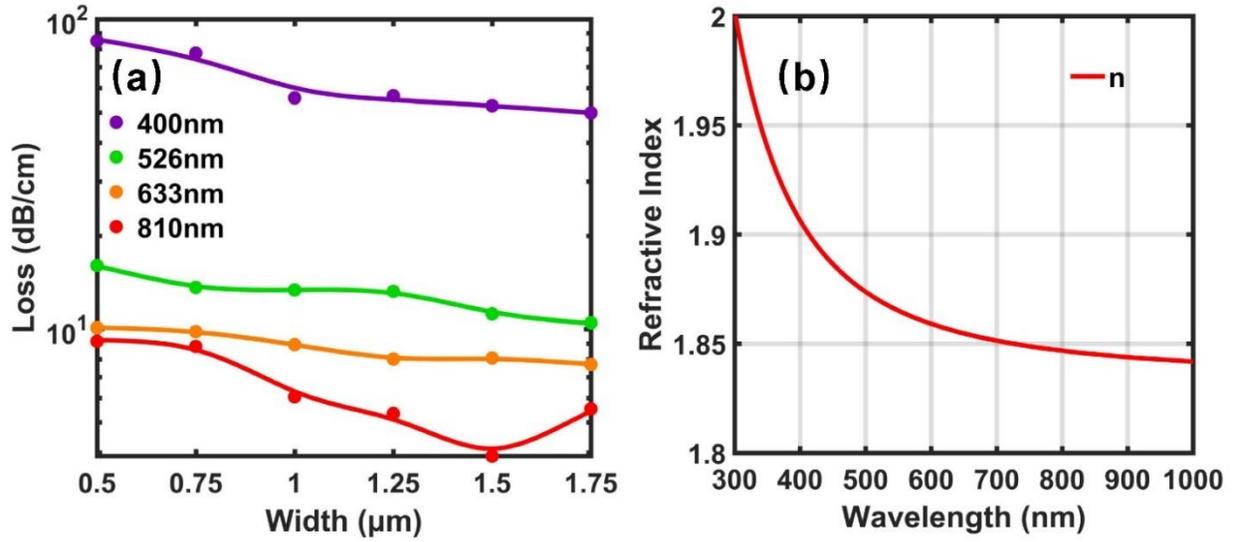

FIG. 3. (a) Propagation losses of different widths and wavelengths. (b) Reflective index n of β-$Ga_2O_3$ films.

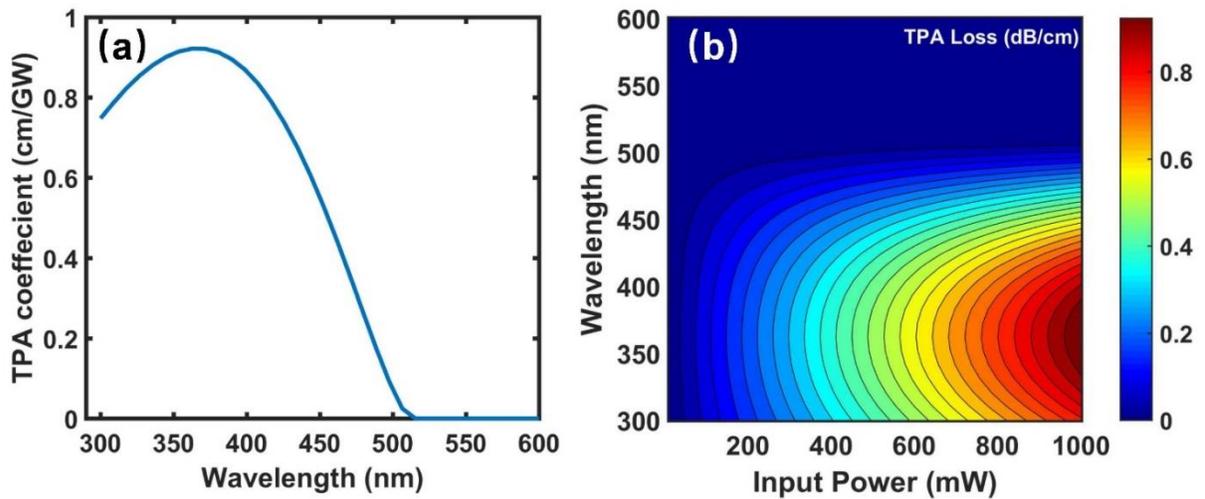

FIG. 4. (a) Theoretical TPA coefficient at different wavelengths. (b) TPA losses at different wavelengths and different input powers.

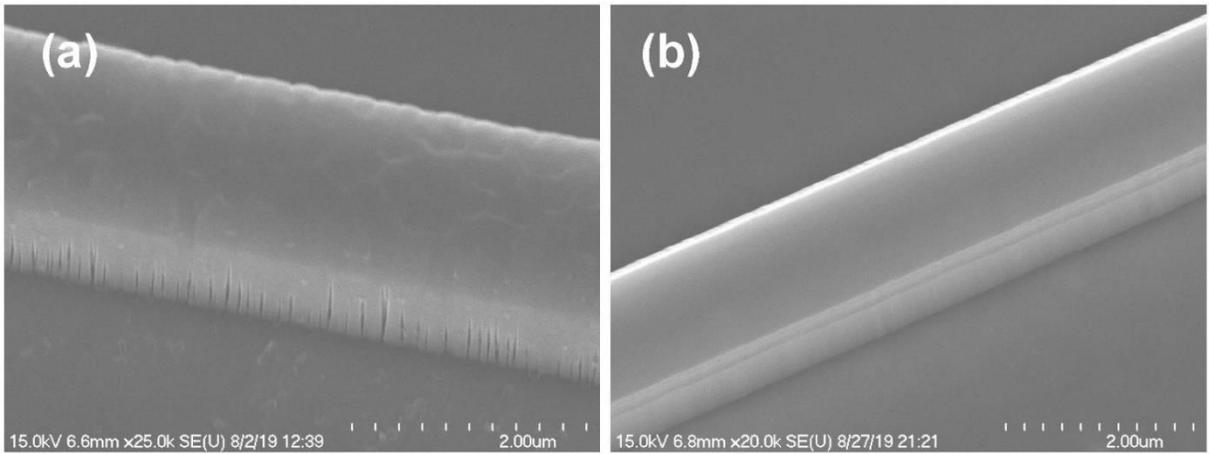

FIG. 5 SEM images of waveguide sidewall. (a) Unoptimized ICP recipe. (b) Optimized ICP recipe.

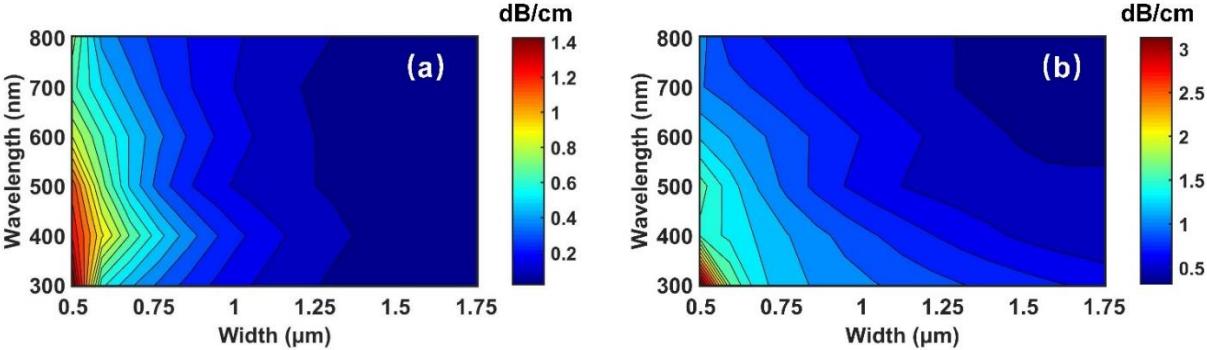

FIG. 6 (a) Theoretical sidewall scattering loss. (b) Theoretical top surface scattering loss.

TABLE I: Unoptimized and Optimized ICP etching recipe.

| Etching Recipe | ICP | Bias | Pressure | BCl$_3$/Ar | DC |
|---|---|---|---|---|---|
| Unoptimized | 800 W | 200 W | 5 mtorr | 50/5 sccm | 290 V |
| Optimized | 400 W | 200 W | 5 mtorr | 20/5 sccm | 330 V |

TABLE II: Contributions of different loss mechanisms to total propagation losses (unit of dB/cm).

| | Total Loss | TPA | Top surface Scattering | Sidewall Scattering | Bulk Scattering |
|---|---|---|---|---|---|
| (1) Width = 1.5 um $\lambda$ = 810 nm | 3.7 | 0 | 0.38 | 0.06 | 3.26 |
| (2) Width = 0.5 um $\lambda$ = 810 nm | 9.14 | 0 | 1.07 | 7.16 | 0.90 |
| (3) Width = 1.5 um $\lambda$ = 526 nm | 11.18 | 0 | 0.49 | 0.60 | 10.09 |